\documentclass[12pt]{article}
\usepackage{epsfig,graphicx}
\usepackage{fpcp04}

\def\beq{\begin{equation}}
\def\eeq#1{\label{#1}\end{equation}}
\def\eeqn{\end{equation}}

\def\beqa{\begin{eqnarray}}
\def\eeqa#1{\label{#1}\end{eqnarray}}
\def\eeqan{\end{eqnarray}}

\let\bar=\overbar

\def\Dslash{\not{\hbox{\kern-4pt $D$}}}
\def\dslash{\not{\hbox{\kern-2pt $\del$}}}

\def\msb{{\bar{\ssstyle M \kern -1pt S}}}

\def\BB0bar{B^0 {\overline B}^0}
\def\BB0dbar{B_d^0 {\overline B}_d^0}
\def\BB0sbar{B_s^0 {\overline B}_s^0}

\RequirePackage{xspace}

\usepackage{relsize}
\def\babar{\mbox{\slshape B\kern-0.1em{\smaller A}\kern-0.1em
    B\kern-0.1em{\smaller A\kern-0.2em R}}}

\def\Kbar  {\kern 0.2em\overline{\kern -0.2em K}{}\xspace}

\def\Kz    {\ensuremath{K^0}\xspace}
\def\Kzb   {\ensuremath{\Kbar^0}\xspace}
\def\KzKzb {\ensuremath{\Kz \kern -0.16em \Kzb}\xspace}
\def\Kp    {\ensuremath{K^+}\xspace}
\def\Km    {\ensuremath{K^-}\xspace}

\def\KpKm  {\ensuremath{\Kp \kern -0.16em \Km}\xspace}

\def\Dbar    {\kern 0.2em\overline{\kern -0.2em D}{}\xspace}

\def\Dz      {\ensuremath{D^0}\xspace}
\def\Dzb     {\ensuremath{\Dbar^0}\xspace}
\def\DzDzb   {\ensuremath{\Dz {\kern -0.16em \Dzb}}\xspace}
\def\Dp      {\ensuremath{D^+}\xspace}
\def\Dm      {\ensuremath{D^-}\xspace}

\def\DpDm    {\ensuremath{\Dp {\kern -0.16em \Dm}}\xspace}

\def\Bbar    {\kern 0.18em\overline{\kern -0.18em B}{}\xspace}

\def\BB      {\ensuremath{B\Bbar}\xspace} 
\def\Bz      {\ensuremath{B^0}\xspace}
\def\Bzb     {\ensuremath{\Bbar^0}\xspace}
\def\BzBzb   {\ensuremath{\Bz {\kern -0.16em \Bzb}}\xspace}
\def\Bu      {\ensuremath{B^+}\xspace}
\def\Bub     {\ensuremath{B^-}\xspace}

\def\BpBm    {\ensuremath{\Bu {\kern -0.16em \Bub}}\xspace}

\mathchardef\Upsilon="7107
\def\Y#1S{\ensuremath{\Upsilon{(#1S)}}\xspace}

\mathchardef\Deltares="7101
\mathchardef\Xi="7104
\mathchardef\Lambda="7103
\mathchardef\Sigma="7106
\mathchardef\Omega="710A

\def\Deltabar{\kern 0.25em\overline{\kern -0.25em \Deltares}{}\xspace}
\def\Lbar{\kern 0.2em\overline{\kern -0.2em\Lambda\kern 0.05em}\kern-0.05em{}\xspace}
\def\Sigbar{\kern 0.2em\overline{\kern -0.2em \Sigma}{}\xspace}
\def\Xibar{\kern 0.2em\overline{\kern -0.2em \Xi}{}\xspace}
\def\Obar{\kern 0.2em\overline{\kern -0.2em \Omega}{}\xspace}
\def\Nbar{\kern 0.2em\overline{\kern -0.2em N}{}\xspace}
\def\Xb{\kern 0.2em\overline{\kern -0.2em X}{}\xspace}

\newcommand{\tev}{\ensuremath{\mathrm{\,Te\kern -0.1em V}}\xspace}
\newcommand{\gev}{\ensuremath{\mathrm{\,Ge\kern -0.1em V}}\xspace}
\newcommand{\mev}{\ensuremath{\mathrm{\,Me\kern -0.1em V}}\xspace}
\newcommand{\kev}{\ensuremath{\mathrm{\,ke\kern -0.1em V}}\xspace}
\newcommand{\ev}{\ensuremath{\mathrm{\,e\kern -0.1em V}}\xspace}
\newcommand{\gevc}{\ensuremath{{\mathrm{\,Ge\kern -0.1em V\!/}c}}\xspace}
\newcommand{\mevc}{\ensuremath{{\mathrm{\,Me\kern -0.1em V\!/}c}}\xspace}
\newcommand{\gevcc}{\ensuremath{{\mathrm{\,Ge\kern -0.1em V\!/}c^2}}\xspace}
\newcommand{\mevcc}{\ensuremath{{\mathrm{\,Me\kern -0.1em V\!/}c^2}}\xspace}

\def\mus  {\ensuremath{\rm \,\mus}\xspace}

\def\mus        {\ensuremath{\,\mu{\rm s}}\xspace}    

\def\to                 {\ensuremath{\rightarrow}\xspace}

\def\pep2{PEP-II}

\def\gsim{{~\raise.15em\hbox{$>$}\kern-.85em
          \lower.35em\hbox{$\sim$}~}\xspace}
\def\lsim{{~\raise.15em\hbox{$<$}\kern-.85em
          \lower.35em\hbox{$\sim$}~}\xspace}

\xspace

\newcommand{\lqcd}{\ensuremath{\Lambda_{\mathrm{QCD}}}\xspace}

\def\jetset74   {\mbox{\tt Jetset \hspace{-0.5em}7.\hspace{-0.2em}4}\xspace}

\def\Dslash{D\!\!\!\!\slash}

\def\nslash{n\!\!\!\slash}
\def\bnslash{\bar n\!\!\!\slash}

\def\dslash{\partial\!\!\!\slash}

\def\OMIT#1{}

\newcommand{\nn}{\nonumber} 

\newcommand{\bn}{{\bar n}}
\newcommand{\bea}{\begin{eqnarray}}
\newcommand{\eea}{\end{eqnarray}}

\newcommand{\SCETa}{\mbox{${\rm SCET}_{\rm I}$ }}

\def\lqcd{\Lambda_{\rm QCD}}
\begin{document}

\Title{Theory of hadronic B decays}
\bigskip


\label{PirjolStart}

%
\author{ Dan Pirjol\index{DP} }

%
\address{Center for Theoretical Physics, MIT\\
77 Massachusetts Avenue, Cambridge, MA 02139\\
}

\makeauthor\abstracts{
I give an overview of the theory of hadronic nonleptonic B decays into
two light mesons. Using the soft-collinear effective theory (SCET), a 
factorization theorem for these processes has been proven to leading
order in $1/m_b$. The phenomenological implications of this
factorization relation for $B\to \pi\pi$ decays are discussed, together
with the prospects for determining $\alpha$ from these modes. 
}

\section{Introduction}

The hadronic decays of B mesons provide a unique source of information
about the flavor structure of the Standard Model. Due to the peculiar
hierarchy structure of the CKM matrix, CP violation is an order unity effect 
in $B$ decays. After several years of data taking from the B factories 
BABAR and BELLE, we are now in a position to perform precision tests of the 
CKM mechanism for CP violation.

It is therefore of considerable interest to have a better theoretical 
understanding of the hadronic dynamics of B decays. Two main approaches are 
widely followed:
a) flavor symmetry methods \cite{GHLR,BuFl}, where isospin or 
SU(3) flavor symmetry are used to reduce the number of independent hadronic amplitudes.
b) dynamical approaches, based on the $1/m_b$ expansion and factorization
theorems. Several such methods have been proposed and used extensively over
the past few years, known as `QCD factorization' (QCDF) 
\cite{BBNS} and `pQCD' \cite{pQCD}. Recently, an effective theory
approach based on the Soft-Collinear Effective Theory (SCET) \cite{SCET} has been
used to study these decays.
The flavor symmetry approach is covered at this conference in the talk of
J. Zupan \cite{zupan-fpcp04} and some aspects of the second approach in the 
talk of H.~Y.~Cheng \cite{cheng-fpcp04}. I will discuss here
recent progress on hadronic decays using the SCET.

\section{SCET factorization relation}

The weak Hamiltonian mediating non-leptonic $B$ decays is given by
\begin{eqnarray} \label{HW}
 H_W = \frac{G_F}{\sqrt{2}}\sum_{f=d,s} \left[ V_{ub} V_{uf}^*
 \left( C_1 O_1^u + C_2 O_2^u  \right) +
V_{cb} V_{cf}^*
 \left( C_1 O_1^c + C_2 O_2^c  \right)
  - V_{tb} V_{tf}^* \sum_{i=3}^{10}\!\! C_i O_i^f \Big) \right],\nn
\end{eqnarray}
where $f=d,s$ for $\Delta S = 0,1$
transitions, respectively.  The tree operators $O_{1,2}^q$ are defined as
\begin{eqnarray}\label{O12}
 O_1^q \!\! &=&\!\! (\bar q b)_{V\!-\!A}
  (\bar f q)_{V\!-\!A}, \ \
 O_2^q = (\bar q_{\beta} b_{\alpha})_{V\!-\!A}
  (\bar f_{\alpha} q_{\beta})_{V\!-\!A}, \nonumber 
\end{eqnarray}
while $O_{3-6}$ are the so-called QCD penguin operators, and $O_{7-10}$
are the electroweak (EW) penguins.
The Hamiltonian Eq.~(\ref{HW}) is matched onto \SCETa
at the scale $Q\sim m_b$ 
\begin{eqnarray} \label{SCETeff}
 H_W  &=  & \frac{2G_F}{\sqrt{2}} \sum_{n,\bn} \bigg\{ 
  \sum_i \int [d\omega_{j}]_{j=1}^{3}
       c_i(\omega_j)  Q_{i}^{(0)}(\omega_j)
  + \sum_i \int [d\omega_{j}]_{j=1}^{4}  b_i(\omega_j) 
  Q_{i}^{(1)}(\omega_j) 
  + {\cal Q}_{c\bar c} + \ldots \bigg\} 
\end{eqnarray}
where $c_i^{(f)}(\omega_j)$ and $b_i^{(f)}(\omega_j)$ are Wilson coefficients, the ellipses
denote color-octet operators which do not contribute at leading order and higher order terms 
in $\lqcd/Q$, $Q=\{m_b,E\}$, and ${\cal Q}_{c\bar c}$
denotes operators containing a $c \bar c$ pair. Their precise form is not required
in the following. We omit the dependence of the Wilson coefficients on the labels 
$\omega_j$. The SCET operators appearing here are defined as ($q=u,d,s$)
\begin{eqnarray} \nn
\begin{array}{cc}
O(\lambda^0) &  O(\lambda) \\
\hline
  Q_{1}^{(0)} =  \big[ \bar u_{n,\omega_1} \bnslash P_L b_v\big]
  \big[ \bar f_{\bn,\omega_2}  \nslash P_L u_{\bn,\omega_3} \big]
  & 
  Q_{1}^{(1)} = \frac{-2}{m_b} 
     \big[ \bar u_{n,\omega_1}\, ig\,\slash\!\!\!\!{\cal B}^\perp_{n,\omega_4} 
     P_L b_v\big]
     \big[ \bar f_{\bn,\omega_2}  \nslash P_L u_{\bn,\omega_3} \big] \\
  Q_{2,3}^{(0)} =  \big[ \bar f_{n,\omega_1} \bnslash P_L b_v \big]
  \big[ \bar u_{\bn,\omega_2} \nslash P_{L,R} u_{\bn,\omega_3} \big]
  &
  Q_{2,3}^{(1)} =  \frac{-2}{m_b}  
     \big[ \bar f_{n,\omega_1} \, ig\,\slash\!\!\!\!{\cal B}^\perp_{n,\omega_4} 
     P_L b_v \big]
     \big[ \bar u_{\bn,\omega_2} \nslash P_{L,R} u_{\bn,\omega_3} \big] \\
  Q_{4}^{(0)} =  \big[ \bar q_{n,\omega_1} \bnslash P_L b_v \big]
  \big[ \bar f_{\bn,\omega_2} \nslash P_{L}\, q_{\bn,\omega_3} \big]
  & 
  Q_{4}^{(1)} =  \frac{-2}{m_b} 
     \big[ \bar q_{n,\omega_1} \, ig\,\slash\!\!\!\!{\cal B}^\perp_{n,\omega_4} 
     P_L b_v \big]
     \big[ \bar f_{\bn,\omega_2} \nslash P_{L}\, q_{\bn,\omega_3} \big] \\
\hline
\end{array} 
\end{eqnarray}
where we have omitted operators which give rise to flavor-singlet light mesons. 
The operators $Q_3$ receive contributions only from electroweak penguins. 
The effective theory operators contain collinear fields along both $n$ and $\bn$
directions \cite{nbn}.

It is convenient to write the Wilson coefficients of the SCET operators 
in a form which separates the contributions from different operators in the 
full theory Hamiltonian
\begin{eqnarray}
c_i &=& \lambda_u^{(f)} c_{iu} + 
\lambda_t^{(f)} \left[c_{it}^{\rm p} + c_{it}^{\rm ew} \right] \,,\qquad
b_i = \lambda_u^{(f)} c_{iu} + 
\lambda_t^{(f)} \left[b_{it}^{\rm p} + b_{it}^{\rm ew} \right] \,.
\end{eqnarray}
The Wilson coefficients of the $O(\lambda^0)$ operators are known to 
$O(\alpha_s(Q))$ \cite{BBNS}, but those of the $O(\lambda^0)$ operators only to 
tree level. The dominant contributions of the EWPs to the SCET
Wilson coefficients
$c_{it}^{\rm ew}$ and $b_{it}^{\rm ew}$ come from $Q_{9,10}$ and
are fixed by SU(3) symmetry to all orders in terms of the coefficients 
of the tree operators $c_{1,2u}$ and $b_{1,2u}$ \cite{BaPi}.

The nonleptonic B decay amplitudes are obtained by taking the
matrix elements of the SCET effective Hamiltonian Eq.~(\ref{SCETeff})
with subleading terms in the usoft-collinear Lagrangian \cite{position}.
The procedure is completely analogous to that followed in deriving
factorization relations for the heavy-to-light form factors \cite{Bauer:2002aj}.
The main result for the $\bar B\to M_n M_\bn$ nonleptonic decay amplitude 
at leading order in $\Lambda/m_b$ can be written in a schematic form as
\begin{eqnarray}\label{fact}
A = c(u) \star \phi_\bn(u) \zeta^{BM_n} + b(x,z,u) \star \phi_n(x) \star \phi_\bn(u) 
\star J(x,z,k_+) \star \phi_B(k_+) + (n \leftrightarrow \bn) +
\langle {\cal Q}_{c\bar c}\rangle
\end{eqnarray}
with $c,b$ Wilson coefficients, $J(x,z,k_+)$ a jet function, $\zeta^{BM_n}$ is a 
nonperturbative soft function and $\phi_n(x), \phi_\bn(u), 
\phi_B(k_+)$ are light-cone wavefunctions for the light mesons and the B meson respectively. 
The corrections to this formula are suppressed by one power of $\Lambda/m_b$.

The main features of this factorization formula are:

\begin{itemize}

\item The soft function $\zeta^{BM}$ is the same as that appearing in the
heavy-to-light form factor $B\to M$ at large recoil \cite{Bauer:2002aj}.

\item Jet universality. The jet function is the same as that 
entering the factorization relation for $B\to P, V_\parallel$
form factors.

\end{itemize}

These points show an unexpected relation between semileptonic 
and nonleptonic decays. This connection can be made more transparent
by defining a new nonperturbative amplitude $\zeta_J^{BM_n}(u,z) \equiv
\phi_n(x) \star J(x,z,k_+) \star \phi_B(k_+)$, which has the same
scaling in $1/m_b$ as $\zeta^{BM_n}$. In the following we will take as independent
nonperturbative parameters $\zeta,\zeta_J(u,z)$, which effectively 
includes perturbative corrections at the collinear scale $\mu^2= Q\Lambda$
to all orders. 

\section{$B\to \pi\pi$ decays}

As an application of the formalism described above we discuss the nonleptonic
$B\to \pi\pi$ decays. The amplitudes can be written 
in a compact form  as
\begin{eqnarray}\label{Apipi}
 A(\bar B^0\to \pi^+\pi^- ) &=& 
\lambda_u^{(d)} (-T-P_{u}) + \lambda_c^{(d)}(-P_c) + \lambda_t^{(d)} (-P_{t})
\equiv
\lambda_u^{(d)} T_c ( 1 + r_c e^{i\delta_c} e^{i\phi}) \\
\sqrt2 A(\bar B^0\to \pi^0 \pi^0) &=& 
\lambda_u^{(d)} (-C +P_u) +  \lambda_c^{(d)} P_c +
\lambda_t^{(d)} P_t \equiv
\lambda_u^{(d)} T_n (1 + r_n e^{i\delta_n} e^{i\phi})\nonumber \\
\sqrt2 A(B^-\to \pi^- \pi^0) &=&
\lambda_u^{(d)} (T + C) \nn
\end{eqnarray}
We neglected here small contributions from electroweak penguins, which can be
included in a model-independent way using isospin symmetry. The amplitudes
on the rhs are defined as $T_c = - T - P_u + P_t, T_n = -C+P_u-P_t$ and
$\phi = \gamma$.

The $B\to \pi\pi$ data is shown in Table 1 \cite{pipiexp}.
This includes the branching ratios and the time-dependent CP violation
parameters $S_{\pi\pi}, C_{\pi\pi}$ in $B^0(t)\to \pi^+\pi^-$. 
The relevant branching ratio information is contained in the two ratios
\begin{eqnarray}
R_c = \frac{Br(B^0\to \pi^+\pi^-)}{2 Br(B^-\to \pi^0\pi^-)}\frac{\tau_{B^+}}{\tau^{B^0}}
= 0.445 \pm 0.062\,,\quad
R_n = \frac{Br(B^0\to \pi^0\pi^0)}{Br(B^-\to \pi^0\pi^-)}\frac{\tau_{B^+}}{\tau^{B^0}}
= 0.292 \pm 0.063
\end{eqnarray}
We show in Table I also $C_{\pi^0\pi^0}$, the direct CP asymmetry
in $B^0\to \pi^0\pi^0$, which was recently measured by the Babar and BELLE Collaborations.

\subsection{Isospin analysis}

For a given $\gamma$, the data on $S_{\pi\pi}, C_{\pi\pi}, R_c, R_n$
allows the determination
of the amplitude parameters $T,C,P$ in Eqs.~(\ref{Apipi}). 
Adding in also $C_{\pi^0\pi^0}$, the weak phase $\alpha = \pi -\beta-\gamma$ 
can be determined with a four-fold ambiguity. This is the well-known isospin
analysis of Gronau and London \cite{GL}. We present first the isospin
analysis for fixed $\gamma$, and then compare
the results with the SCET predictions. We discuss the 
prospects for a $\gamma$ (or $\alpha$) determination in Sec.~4.

We will present the results of the isospin analysis in terms of the
parameters $(r_c, \delta_c, u, v)$, where $t = (u,v)$ are the coordinates
of the apex of the
triangle of isospin amplitudes $1 + t_n = t$. We defined here
$t = T/T_c, t_n = T_n/T_c$. The measurable parameters are given by
\begin{eqnarray}
S_{\pi\pi} &=& 
\frac{-\sin(2\beta+2\gamma) - 2r_c \cos\delta_c \sin(2\beta+\gamma)-
r_c^2\sin 2\beta}{1+2r_c \cos\delta_c \cos\gamma + r_c^2}\\
C_{\pi\pi} &=& 
\frac{2r_c \sin\delta_c \sin\gamma }
{1+2r_c \cos\delta_c \cos\gamma + r_c^2}\\
R_c &=& \frac{1}{t^2}[1 + r_c^2 + 2r_c \cos\delta_c \cos\gamma]\,,\quad
R_n = \frac{1}{t^2}[t_n^2 + r_c^2 - 2
r_c (\cos\delta_c (u-1) + \sin\delta_c v) \cos\gamma]
\end{eqnarray}
with $t^2 = u^2+v^2, t_n^2=(u-1)^2+v^2$. The direct CP asymmetry in the
$B^0\to \pi^0\pi^0$ mode is
\begin{eqnarray}
C_{\pi^0\pi^0} = -
\frac{2[r_c \sin\delta_c(u-1) - r_c\cos\delta_c v] \sin\gamma }
{t_n^2 - 2[r_c \cos\delta_c (u-1) + r_c \sin\delta_c v] \cos\gamma + r_c^2}
\end{eqnarray}

We show in Table 2 the results for the amplitude parameters extracted 
from the data corresponding to several values of $\gamma$.
For each given value of $\gamma$, there are four solutions for the
parameters $(r_c,\delta_c,u,v)$, which fall into two sets with common
values of $(r_c,\delta_c)$. We select only the physical solution corresponding
to $r_c \leq 1$, which gives the 2 solutions for the 
amplitudes $T_c, T_n$ shown in Table 2.
For each of these solutions we show also predictions for the
direct CP asymmetry in the neutral pions channel $C_{\pi^0\pi^0}$.
Similar analyses have been presented in \cite{BuFl,pheno}.

\begin{table}
\begin{center}
\scriptsize
\begin{tabular}{|c|ccc|} 
\hline
 & BABAR & Belle & Avg. \\
\hline
$Br(B^+\to \pi^+ \pi^0)$                    & 
$5.8\pm 0.6\pm 0.4$                         & 
$5.0\pm 1.2\pm 0.5$                         & 
$5.61\pm 0.61$                              \\ 

$Br(B^0\to \pi^0 \pi^0)$                    & 
$1.17\pm 0.32 \pm 0.10$                     & 
$2.32^{+0.44+0.22}_{-0.48-0.18}$            & 
$1.51\pm 0.28$                              \\ 

$Br(B^+\to \pi^+ \pi^-)$                    & 
$4.7\pm 0.6\pm 0.2$                         & 
$4.4\pm 0.6 \pm 0.3$                        & 
$4.6\pm 0.4$                                \\ 

\hline

$S_{\pi \pi}$                               & 
$-0.30\pm 0.17\pm 0.03$                     & 
$-1.00\pm 0.22$                             & 
$-0.608\pm 0.135$                           \\ 

$C_{\pi \pi}$                               & 
$-0.09\pm 0.15\pm 0.04$                     & 
$-0.58\pm 0.17$                             & 
$-0.368\pm 0.112$                           \\ 
\hline

$C_{\pi^0 \pi^0}$                           & 
$-0.12\pm 0.56 \pm 0.06$                    & 
$-0.43\pm 0.51^{0.16}_{-0.17}$              & 
$-0.28\pm 0.39$                             \\ 
\hline
\end{tabular}
\end{center}
{\caption{Data on $B\to \pi\pi$ decays after ICHEP 2004,  Beijing \cite{HFAG,pipiexp}. 
The CP-averaged branching ratios are
quoted in units of $10^{-6}$. }}
\end{table}

\begin{table}
\begin{center}
\scriptsize

\begin{tabular}{|c|c|c||c|}
\hline
\hline
$\gamma$ & $(r_c,\delta_c)$ & $(u,v)$ & 
$C_{\pi^0\pi^0}$ \\
\hline
$54^\circ$ & $(0.32\pm 0.11\,, -1.12\pm 0.40)$ &  
\begin{tabular}{c}
$(1.36\pm 0.21\,, -1.00\pm 0.17)$ \\ 
$(1.60\pm 0.22\,, 0.53\pm 0.21)$ \\
\end{tabular}
 & 
\begin{tabular}{c}
$-0.07\pm 0.29$ \\
$0.48\pm 0.21$ \\
\end{tabular} \\
\hline
$64^\circ$ & $(0.49\pm 0.14\,, -0.71\pm 0.27)$ &  
\begin{tabular}{c}    
$(1.63\pm 0.27\,, -0.92\pm 0.21)$ \\
$(1.80\pm 0.27\,, 0.51\pm 0.30)$  \\
\end{tabular}
&
\begin{tabular}{c}
$-0.24\pm 0.34$ \\
$0.78\pm 0.21$ \\
\end{tabular} \\
\hline
$74^\circ$ & $(0.68\pm 0.15\,, -0.53\pm 0.20)$  & 
\begin{tabular}{c}
$(1.95\pm 0.41\,, -0.43\pm 1.03)$ \\
$(2.00\pm 0.27\,, 0.11\pm 1.12)$ \\
\end{tabular}
  & 
\begin{tabular}{c}
$0.12\pm 1.10$ \\ 
$0.67\pm 0.84$ \\
\end{tabular} \\
\hline
\hline
\end{tabular}
\end{center}
{\caption{Amplitude parameters in $B\to \pi\pi$ for several
input values for $\gamma$, together with the prediction for the
CP asymmetry in $B\to \pi^0\pi^0$. For each value of $\gamma$ there are
two solutions for the tree amplitudes.}
\label{amps:pipi} }
\end{table}

The absolute magnitudes of the amplitudes are set by
$|T_c+T_n| = N_\pi (0.296 \pm 0.016)$ GeV, which can be extracted from 
Br$(B^- \to \pi^-\pi^0)$. [We denoted here
$N_\pi = G_F/\sqrt2 m_B^2 f_\pi$ and used $|V_{ub}|=0.0039$.]

\subsection{SCET analysis}

The analysis discussed above used only isospin symmetry.
Next we examine the predictions from the SCET factorization formula.

1. {\em Predicting $Br(B^0\to \pi^0\pi^0)$.}
At tree level in matching, the strong phases of the tree amplitudes
vanish Arg$(T_n/T_c) \sim O(\alpha_s(Q), \Lambda/Q)$. This fixes
one hadronic parameter $(v\to 0)$, such that 
$R_c, R_n$ and $C_{\pi^0\pi^0}$ are not independent quantities, but are related as
\begin{eqnarray}\label{RnSCET}
R_n = \frac{1}{t}[(t-1)(1 - t R_c) + r_c^2]\,,\qquad
C_{\pi^0\pi^0}  = - (t-1)\frac{R_c}{R_n} C_{\pi^+\pi^-}
\end{eqnarray}
This allows predictions for $R_n(\alpha)$ and $C_{\pi^0\pi^0}(\alpha)$ to be made
using only data on $R_c, S_{\pi\pi}, C_{\pi\pi}$. We show in Fig. 1 (a) the 
prediction for $R_n(\alpha)$ from Eq.~(\ref{RnSCET}) as a function of $\alpha$.

2. {\em Determining the SCET nonperturbative parameters.}
The LO factorization relation for $T$ and $T_c$ expresses these amplitudes
in terms of SCET Wilson coefficients and the 
nonperturbative parameters $\zeta, \zeta_J(x)$. 
Working at tree level in matching, one finds
\begin{eqnarray}\label{SCETLO}
\left. \zeta^{B \pi} \right|_{\gamma = 64^\circ} =
(0.08 \pm 0.03) \left( \frac{3.9 \times 10^{-3}}{|V_{ub}|} \right) \,,\qquad
\left. \zeta_J^{B \pi} \right|_{\gamma = 64^\circ} =
(0.10 \pm 0.02) \left( \frac{3.9 \times 10^{-3}}{|V_{ub}|} \right) \,,
\end{eqnarray}
which does not include any theoretical uncertainties.
These values can be used to predict the $B\to \pi$ form factor $f_+(0)$
at $q^2=0$ as
\begin{eqnarray}
&&f_+(0)= \zeta^{B\pi} + \zeta_J^{B\pi} = \nn
(0.18 \pm 0.05) \left( \frac{3.9 \times 10^{-3}}{|V_{ub}|} \right) 
\end{eqnarray}
which is somewhat lower than recents results from QCD sum rules
$f_+(0)= 0.26\pm 0.03$ \cite{QCDSR}. 
The result for $\zeta_J^{B\pi}$ in Eq.~(\ref{SCETLO}) is slightly higher than
the values obtained in perturbation theory working at tree level in the
jet function in Eq.~(\ref{fact}) 
\begin{eqnarray}
\zeta_J^{B\pi} = \frac{\pi\alpha_s C_F}{N_c} \frac{f_B f_\pi}{m_B}
\langle \bar x^{-1}\rangle_\pi
\langle k_+^{-1}\rangle_B \sim 0.02 - 0.05
\end{eqnarray}
where we took $ \langle \bar x^{-1}\rangle_\pi=
3(1+a_2^\pi)$ with $a_2^\pi = 0.2\pm 0.2$, $f_B = 200$  MeV, $f_\pi =131$ MeV
and $\langle k_+^{-1}\rangle_B = 1/\lambda_B$ with 
$\lambda_B =(350\pm 150)$ MeV \cite{BBNS}. [The $O(\alpha_s^2(m_b \Lambda))$
corrections to this result have been recently obtained in \cite{jetoneloop}.]
Conceivable explanations for this discrepancy 
are experimental errors in the $B\to \pi\pi$ data, or neglected power corrections
to the factorization formula.
A detailed analysis using the QCDF approach
\cite{comparison} shows that power corrections are small in the tree amplitudes 
$T,T_c$ and thus do not affect significantly this determination of 
$\zeta_J^{B\pi}$. 

\section{Prospects for determining $\alpha$ }

The main motivation for the experimental study of the $\Delta S = 0$ decays
such as $B\to \pi^+\pi^-$ is
in connection with the determination of the angle $\alpha$. 
In fact what is measured is the combination $\beta+\gamma = \pi - \alpha$,
which taken together with the precise value of $\beta$ known from
charmonium modes, can be translated into a value of $\gamma$.

The measurements are usually expressed in terms of an effective
angle $\alpha_{\rm eff}$ defined by $\sin 2\alpha_{\rm eff} =
S_{\pi\pi}/\sqrt{1-C^2_{\pi\pi}}$. This is related to the physical
angle by $\alpha_{\rm eff} = \alpha - \theta$ with 
$\theta = Arg(A_{\pi^+\pi^-}/\bar A_{\pi^+\pi^-})$. Using only data on 
$R_c,R_n,C$, only bounds on $\theta$ can be obtained. 
These bounds can be turned into a determination provided that 
$C_{\pi^0\pi^0}$ is also measured \cite{GL}.

Several  bounds on $\theta$ using only isospin have been given in 
Refs.~\cite{bounds,GLSS}, of which the most restrictive one is
the GLSS bound \cite{GLSS}
\begin{eqnarray}\label{glss}
\cos 2\theta \geq \frac{1}{2R_c \sqrt{1-C^2_{\pi\pi}}}
[(R_c+1-R_n)^2-2R_c]
\end{eqnarray}
With the present data in Table I these bounds constrain $\alpha$
to lie within 4 windows of width $2\theta$
with $\theta = 29.0^\circ$
centered on $\alpha_{\rm eff} = 1/2 \arcsin(S_{\pi\pi}/\sqrt{1-C^2_{\pi\pi}})$. 
The window corresponding to the physically preferred solution has
$\alpha_{\rm eff}^{\rm I} = 110.41^\circ$.
These can be translated into bounds on $\beta+\gamma$ 
which gives $40.6^\circ = 69.6^\circ - \theta \leq
\gamma + \beta \leq 69.6^\circ + \theta = 98.6^\circ$.

These bounds can be turned into a determination of $\alpha$ provided that
$C_{\pi^0\pi^0}$ is known, which is equivalent to performing the full
isospin analysis. In the absence of this information, one can restrict
the ranges of the bound by adding in dynamical information about the
amplitudes $T,P,C$. Several such ``constrained'' bounds exist, of which 
we mention only two.

\begin{itemize}

\item The Buchalla-Safir bound \cite{BuSa}. This assumes that
$|\delta_c | \leq \pi/2$, which leads to a lower bound on $\bar \eta$
as a function of $S_{\pi\pi}$.

\item The SCET constraint \cite{Bauer:2004tj,scetgamma}. The dynamical
input here is the smallness of the relative strong phase of $T_c$ and $T_n$.
\begin{eqnarray}\label{corr}
Arg(T/T_c) \sim Arg(T_n/T_c) \sim O(\alpha_s(m_b), \Lambda/m_b)
\end{eqnarray}
\end{itemize}

\begin{figure}[t!]
 \centerline{
\epsfysize=6truecm \hbox{\epsfbox{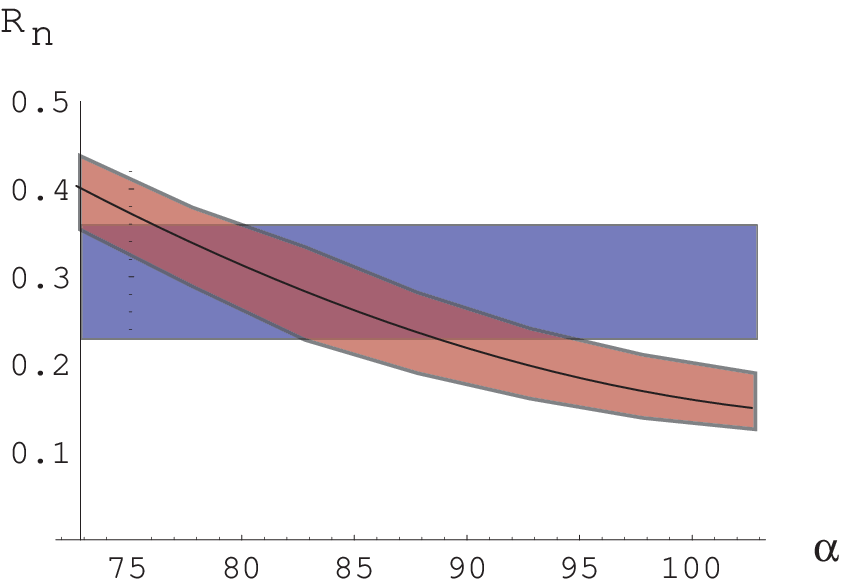}} 
\raisebox{-1cm}
{\epsfysize=7truecm \hbox{\epsfbox{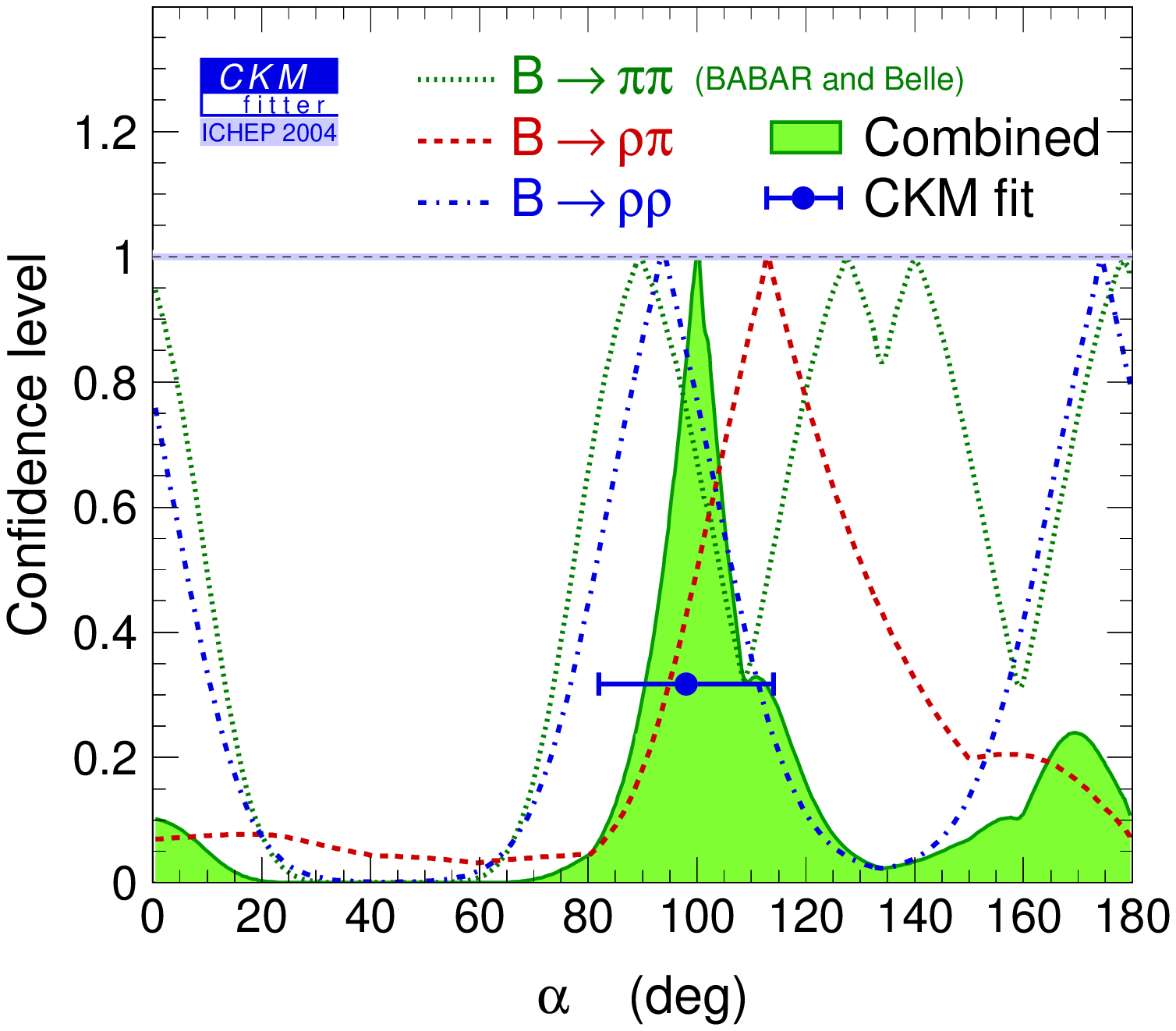}}}  
} 
\hspace{3cm}(a) \hspace{8cm} (b)
{\caption[1]{(a) Constraint on the weak phase $\alpha$ 
following from a small relative strong phase between the tree amplitudes $T,T_c$. 
The light band shows the $1\sigma$ prediction for $R_n(\alpha)$ as a function
of the weak phase $\alpha$ following from Eq.~(\ref{RnSCET}); the horizontal 
band
shows the measured value $R_n=0.292\pm 0.063$. 
The solid line denotes the GLSS bound 
$R_n \geq R^{GLSS}_n(\alpha, R_c, S_{\pi\pi}, C_{\pi\pi})$ for central values
of the parameters.
(b) constraints on $\alpha$ from charmless $B$ decays \cite{ZICHEP}.}
\label{fig:CKMf} }
\vskip -0.3cm
\end{figure}

We will discuss here in some detail the SCET constraint.
We show in Fig.~1 (a) the constraints on $\alpha$ from comparing
the prediction for $R_n(\alpha)$ from requiring a flat tree triangle Eq.~(\ref{RnSCET})
with the measured value of this parameter $R_n = 0.292 \pm 0.063$. One additional
constraint is introduced by the GLSS bound (\ref{glss}) which bounds
$R_n$ from below $R_n \geq R_n^{GLSS}(\alpha,R_c,C_{\pi\pi})$.
 Taking into 
account all these constraints, the plot in Fig.~1 (a) allows the range for $\alpha$
\begin{eqnarray}
73^\circ \leq \alpha \leq 95^\circ
\end{eqnarray}
which includes only the experimental uncertainties. This result is in 
agreement with present constraints on $\alpha$ from $\Delta S=0$
modes (see Fig.~1(b)) and with a general constraint combining
$B\to \pi\pi, \rho\pi, \rho\rho$ modes \cite{alphacons}.

A method for determining $\gamma$ based on this constraint was proposed in
Ref.~\cite{scetgamma}, where it was argued that theoretical 
uncertainties to the 
condition Eq.~(\ref{corr}) from radiative corrections and power corrections 
of canonical size introduce a very small theoretical error on $\alpha$ of 
$\pm 2^\circ$ (for the present central values of the data). 
More detailed theoretical computations of the correction to Eq.~(\ref{corr}) 
would be welcome. 
With improved data the SCET constraint Eq.~(\ref{corr}) can 
be expected to give useful information on $\alpha$, complementing alternative 
determinations of this weak phase.

\subsection{Acknowledgments}
I would like to thank the organizers for an enjoyable conference. 
This work has been supported by the U.S.\ Department of Energy under cooperative
research agreement 
DOE-FC02-94ER40818.

%
\label{PirjolEnd}


\begin{thebibliography}{99}

\bibitem{GHLR}
M.~Gronau, O.~F.~Hernandez, D.~London and J.~L.~Rosner,
Phys.\ Rev.\ D {\bf 50}, 4529 (1994);
Phys.\ Rev.\ D {\bf 52}, 6374 (1995).

\bibitem{BuFl}
A.~J.~Buras, R.~Fleischer, S.~Recksiegel and F.~Schwab,
arXiv:hep-ph/0402112;
A.~J.~Buras, R.~Fleischer, S.~Recksiegel and F.~Schwab,
Eur.\ Phys.\ J.\ C {\bf 32}, 45 (2003);

\bibitem{zupan-fpcp04}
J. Zupan, {\em Determining $\alpha$ and $\gamma$ - theory},
[arXiv:hep-ph/0410371].


\bibitem{cheng-fpcp04} 
H.~Y.~Cheng,
{\em Direct CP violation and final state interactions in hadronic B decays},
[arXiv:hep-ph/0411340].


\bibitem{BBNS}
M.~Beneke, G.~Buchalla, M.~Neubert and C.~T.~Sachrajda,
Phys.\ Rev.\ Lett.\  {\bf 83}, 1914 (1999);
Nucl.\ Phys.\ B {\bf 606}, 245 (2001);
M.~Beneke and M.~Neubert,
Nucl.\ Phys.\ B {\bf 675}, 333 (2003).

\bibitem{pQCD}
Y.~Y.~Keum, H.~N.~Li and A.~I.~Sanda,
Phys.\ Rev.\ D {\bf 63}, 054008 (2001);
C.~H.~Chen, Y.~Y.~Keum and H.~n.~Li,
Phys.\ Rev.\ D {\bf 64}, 112002 (2001);
H.~Y.~Cheng, C.~K.~Chua and A.~Soni,
arXiv:hep-ph/0409317.



\bibitem{SCET}
C.~W.~Bauer, S.~Fleming and M.~E.~Luke,
Phys.\ Rev.\ D {\bf 63}, 014006 (2001);
C.~W.~Bauer, S.~Fleming, D.~Pirjol and I.~W.~Stewart,
Phys.\ Rev.\ D {\bf 63}, 114020 (2001);
C.~W.~Bauer and I.~W.~Stewart,
Phys.\ Lett.\ B {\bf 516}, 134 (2001);
Phys.\ Rev.\ D {\bf 65}, 054022 (2002).

\bibitem{nbn}
C.~W.~Bauer, S.~Fleming, D.~Pirjol, I.~Z.~Rothstein and I.~W.~Stewart,
Phys.\ Rev.\ D {\bf 66}, 014017 (2002).


\bibitem{Chay:2003zp}
J.~g.~Chay and C.~Kim,
Phys.\ Rev.\ D {\bf 68}, 071502 (2003);
Nucl.\ Phys.\ B {\bf 680}, 302 (2004).


\bibitem{Bauer:2002aj}
C.~W.~Bauer, D.~Pirjol and I.~W.~Stewart,
Phys.\ Rev.\ D {\bf 67}, 071502 (2003);
D.~Pirjol and I.~W.~Stewart,
Phys.\ Rev.\ D {\bf 67}, 094005 (2003)
[Erratum-ibid.\ D {\bf 69}, 019903 (2004)];
D.~Pirjol and I.~W.~Stewart,
eConf {\bf C030603}, MEC04 (2003)
[arXiv:hep-ph/0309053].

\bibitem{Bauer:2004tj}
C.~W.~Bauer, D.~Pirjol, I.~Z.~Rothstein and I.~W.~Stewart,
Phys.\ Rev.\ D {\bf 70}, 054015 (2004).


\bibitem{BaPi}
C.~W.~Bauer and D.~Pirjol,
Phys.\ Lett.\ B {\bf 604}, 183 (2004)



\bibitem{position}
M.~Beneke, A.~P.~Chapovsky, M.~Diehl and T.~Feldmann,
Nucl.\ Phys.\ B {\bf 643}, 431 (2002);
M.~Beneke and T.~Feldmann,
Phys.\ Lett.\ B {\bf 553}, 267 (2003)

\bibitem{HFAG} 
The Heavy Flavor Averaging Group, 
http://www.slac.stanford.edu/xorg/hfag/

\bibitem{pipiexp}
B.~Aubert  [BABAR Collaboration],
arXiv:hep-ex/0412037;
K.~Abe {\it et al.}  [Belle Collaboration],
arXiv:hep-ex/0408101.
B.~Aubert  [BABAR Collaboration],
arXiv:hep-ex/0501071.


\bibitem{EWP} M. Neubert and J.L. Rosner,
Phys. Lett. B {\bf 441}, 403 (1998);
A.~J.~Buras and R.~Fleischer, Eur. Phys. J. C {\bf 11}, 93 (1999).
M.~Gronau, D.~Pirjol and T.~M.~Yan,
Phys.\ Rev.\ D {\bf 60}, 034021 (1999);
[Erratum-ibid.\ D {\bf 69}, 119901 (2004)].

\bibitem{GL}
M.~Gronau and D.~London,
Phys.\ Rev.\ Lett.\  {\bf 65}, 3381 (1990).

\bibitem{pheno}
C.~W.~Chiang, M.~Gronau, Z.~Luo, J.~L.~Rosner and D.~A.~Suprun,
Phys.\ Rev.\ D {\bf 69}, 034001 (2004);
(ibid) arXiv:hep-ph/0404073;
A.~Ali, E.~Lunghi and A.~Y.~Parkhomenko,
Eur.\ Phys.\ J.\ C {\bf 36}, 183 (2004);
J.~Charles {\it et al.}  [CKMfitter Group Collaboration],
arXiv:hep-ph/0406184.

\bibitem{jetoneloop}
R.~J.~Hill, T.~Becher, S.~J.~Lee and M.~Neubert,
JHEP {\bf 0407}, 081 (2004)

\bibitem{comparison}
C.~W.~Bauer, D.~Pirjol, I.~Z.~Rothstein and I.~W.~Stewart,
arXiv:hep-ph/0502094.

\bibitem{bounds}
Y.~Grossman and H.~R.~Quinn,
Phys.\ Rev.\ D {\bf 58}, 017504 (1998);
J.~Charles,
Phys.\ Rev.\ D {\bf 59}, 054007 (1999);

\bibitem{GLSS}
M.~Gronau, D.~London, N.~Sinha and R.~Sinha,
Phys.\ Lett.\ B {\bf 514}, 315 (2001).

\bibitem{su3bounds}
D.~Pirjol,
Phys.\ Rev.\ D {\bf 60}, 054020 (1999);
R.~Fleischer,
Phys.\ Lett.\ B {\bf 459}, 306 (1999).

\bibitem{BuSa}
G.~Buchalla and A.~S.~Safir,
Phys.\ Rev.\ Lett.\  {\bf 93}, 021801 (2004).

\bibitem{QCDSR}
P.~Ball and R.~Zwicky,
JHEP {\bf 0110}, 019 (2001);
P.~Ball and R.~Zwicky,
arXiv:hep-ph/0406232.

\bibitem{ZICHEP}
Z.~Ligeti,
arXiv:hep-ph/0408267.


\bibitem{scetgamma}
C.~W.~Bauer, I.~Z.~Rothstein and I.~W.~Stewart,
arXiv:hep-ph/0412120.

\bibitem{alphacons}
M.~Gronau, E.~Lunghi and D.~Wyler,
Phys.\ Lett.\ B {\bf 606}, 95 (2005)
[arXiv:hep-ph/0410170].


\end{thebibliography}
\end{document}